\documentclass[sigconf]{acmart}

\usepackage[linesnumbered, ruled, vlined]{algorithm2e}
\usepackage{amsmath}
\usepackage{amsthm}
\usepackage{booktabs}
\usepackage{diagbox}
\usepackage{fontawesome5}
\usepackage{graphicx}
\usepackage{multirow}
\usepackage{tikz}

\usepackage{graphicx}
\usepackage{multirow}
\usepackage{listings}
\usepackage[justification=centering]{caption}
\usepackage{subcaption}
\usepackage[table]{xcolor}
\usepackage{array}
\usepackage{balance}
\usepackage{fancyvrb}
\usepackage{tabularx}
\usepackage{varwidth}

\usepackage[inline]{enumitem}
\usepackage{enumitem}
\usepackage[referable]{threeparttablex}
\usepackage[T1]{fontenc}
\usepackage{tcolorbox}
\usepackage[export]{adjustbox}
\usepackage{xfrac}
\usepackage{wrapfig}
\usepackage{mathpartir}
\usepackage{stmaryrd}
\usepackage{array}
\usepackage{siunitx}
\usepackage{threeparttable}
\usepackage{tcolorbox}
\usepackage{pifont}
\usepackage{diagbox}
\usepackage{xcolor}
\usepackage{wrapfig}
\usepackage{orcidlink}
\usepackage{perpage} 
\usepackage{nicematrix}
\MakePerPage{footnote} 

\SetKwProg{Al}{Algorithm}{}{}
\SetAlFnt{\small\rmfamily}
\SetKw{Continue}{continue}
\SetKw{Break}{break}
\SetKw{And}{and}
\SetKw{Or}{or}
\SetKw{Not}{not}
\SetKw{In}{in}
\SetKw{Null}{NULL}
\SetKwInOut{Input}{Input}
\SetKwInOut{Output}{Output}

\definecolor{darkspringgreen}{rgb}{0.09,0.45,0.27}
\definecolor{keywordred}{RGB}{212,60,73}
\definecolor{funcnamepurple}{RGB}{111,66,193}
\definecolor{specialblue}{RGB}{0,92,197}

\usetikzlibrary{positioning, shapes.geometric, arrows.meta, calc, fit, backgrounds}
\definecolor{myblue}{RGB}{0, 102, 204}
\definecolor{myblack}{RGB}{0, 0, 0}
\definecolor{lightgrey}{RGB}{245, 245, 245}

\SetCommentSty{alcommfont}

\newcommand{\tabincell}[2]{\begin{tabular}{@{}#1@{}}#2\end{tabular}}

\newcommand{\flob}{{\textsc{FLOB}}}
\newcommand{\btd}{{\textsc{BTD}}}
\newcommand{\dnd}{{\textsc{DnD}}}
\newcommand{\nnscope}{{\textsc{NeuroScope}}}
\newcommand{\nnshield}{{\textsc{NeuroShield}}}

\newcommand{\mdobf}{{\textsc{ModelObfuscator}}}
\newcommand{\dynamo}{{\textsc{DynaMO}}}

\newcommand{\loki}{{\textsc{Loki}}}
\newcommand{\tigress}{{\textsc{Tigress}}}

\newcommand{\gamba}{{\textsc{GAMBA}}}
\newcommand{\promba}{{\textsc{ProMBA}}}

\newcommand{\xsmir}{{\textsc{XSmir}}}

\AtBeginDocument{%
  }

\newcounter{expcount}
\newcommand{\experiment}[2][]{%
  \par\vspace{0.5em}
  \refstepcounter{expcount}%
  \ifx#1\empty\else\label{#1}\fi
  \noindent\textbf{Experiment \theexpcount: #2.}
}

\theoremstyle{definition}

\theoremstyle{remark}

\theoremstyle{lemma}

\theoremstyle{theorem}

\theoremstyle{corollary}

\setcopyright{none}
\renewcommand\footnotetextcopyrightpermission[1]{}

\begin{document}

\title[Protecting Floating-Point Computation for DNN Binaries with MBA Obfuscation]{Protecting Floating-Point Computation for Deep Neural Network Binaries with Mixed Boolean-Arithmetic Obfuscation}


\author{Yikun Hu \quad Zichen Zhao \quad Peixiang Qin
	\quad Ziyi Zhou \quad Jiaping Gui}
\affiliation{%
	\institution{Shanghai Jiao Tong University}
	\country{China}
}

\author{Yuandao Cai \qquad Wensheng Tang}
\affiliation{%
	\institution{The Hong Kong University of Science and Technology}
	\country{China}
}

\begin{abstract}


	Deep neural networks~(DNNs) have become a foundational component of modern computing systems with a wide range of applications, such as computer vision, edge intelligence, etc.
	For the sake of low latency and data privacy, DNN models are increasingly compiled into executables and deployed on local devices.
	However, that exposes the models to model theft, enabling adversaries to recover proprietary assets via reverse engineering techniques.
	%
	While code obfuscation naturally emerges for protecting executables from reverse engineering, existing schemes are primarily designed for traditional programs, focusing on complex control flow structures and integer-based operations.
	They are fundamentally inappropriate for DNN binaries, which exhibit relatively simple code structures and heavily rely on floating-point computation.
	
	In this paper, we propose \flob, an obfuscation framework that protects floating-point computations in DNN binaries using \textbf{Mixed Boolean-Arithmetic~(MBA)} transformations.
	Our approach lifts floating-point values into a higher-precision binary expansion space and performs MBA transformations at the representation level.
	The computation is carried out entirely in the lifted space, and the final result is then projected back to the target precision, yielding outputs that are \textbf{semantically equivalent under the original floating-point format}, without introducing additional rounding error.
	The experimental results show that \flob\ outperforms the state-of-the-art obfuscation schemes against existing reverse engineering analyzers and deobfuscators,
	while preserving computational correctness without introducing additional numerical error, allowing flexible trade-off between protection and performance.
	Specifically,
	\flob\ reduces the operator recovery rate to 4.51\% on average, achieving a 32.82 percentage-point reduction compared to existing methods.
   
\end{abstract}


%
%

\maketitle
\renewcommand{\shortauthors}{Y. Hu et al.}

\section{Introduction}

Deep neural networks~(DNNs) have become a foundational component of modern computing systems, enabling a wide range of applications such as computer vision~\cite{seifert2017visualizations}, natural language processing~\cite{otter2020survey}, and edge intelligence~\cite{ren2023survey}.
These models often embody significant intellectual property, as their development requires substantial investment in data collection, model design, and large-scale training~\cite{narayanan2021efficient}.
In practice, DNN models are increasingly compiled into executable binaries and deployed locally, which is similar to traditional software distribution, to avoid the latency, communication overhead, and potential privacy leakage associated with remote inference~\cite{tvm,zhu2022roller,ma2020rammer,zheng2020ansor,rotem2018glow,xla}.
However, the deployment paradigm also exposes DNN binaries to adversaries, who can leverage reverse engineering techniques to recover valuable model assets, including architectures and parameters, leading to serious model theft risks~\cite{btd,dnd,neuroscope}.
Therefore, protecting the proprietary assets embedded in DNN binaries has become a critical problem.

To this end, recent research has adopted code obfuscation techniques to prevent DNN binaries from reverse engineering.
On one hand, \mdobf~\cite{modelobfuscator} and \nnshield~\cite{neuroshield} achieve the protection by modifying model representations and structures, such as operator renaming, operator fusion, operator computation reconstruction, leaving the data and operations exposed during the runtime.
Thus, they are vulnerable to information extraction by dynamic analysis.
On the other hand, \dynamo~\cite{dynamo} obfuscates linear operations at the beginning of inference, after which the DNN computation is performed on the protected information during execution, and the results are then recovered at the end.
Although it improves resilience to dynamic analysis, the method can only protect linear operations, which are guaranteed to be invertible for recovery, whereas non-linear operations are pervasive in modern DNNs, e.g.,~\texttt{Sigmoid} and \texttt{Softmax}.
More fundamentally, the two kinds of methods largely focus on transforming program representation and execution structures rather than protecting the underlying numerical computation.
Namely, they offer limited protection against analysis techniques that reason about computation semantics~(e.g.,~black-box deobfuscation~\cite{xsmir}), and do not provide dedicated protection for floating-point operations, which constitute the core of DNN execution.

In this paper, we propose \flob\footnote{\flob: \underline{FL}oating-point computation \underline{OB}fuscator}, an obfuscation framework to protect floating-point computations in deployed DNN binaries.
The key insight of this work is that floating-point computation can be lifted into a higher-precision representation space, where values can be scaled into integers.
The transformation enables bitwise operations over the resulting representation, making mixed boolean-arithmetic~(MBA) applicable while preserving the original numerical semantics.
Specifically, \flob\ first lifts floating-point values into a higher-precision binary representation, applies MBA transformations at the representation level, and performs the entire computation in this transformed domain.
Only the final results are projected back to the target precision.
In this way, the computation preserves the original semantics without introducing additional numerical error.
As a result, \flob\ directly protects floating-point computation and is not restricted to specific operator types.
This enables it to provide robust protection against reverse engineering attacks~\cite{dnd}, dynamic analysis~\cite{neuroscope,btd}, and black-box deobfuscation~\cite{xsmir}.

We evaluate \flob\ on seven real-world DNN models, spanning five CV and two NLP models.
The results show that it effectively increases the difficulty for existing reverse engineering and deobfuscation techniques to extract sensitive information from DNN binaries, outperforming state-of-the-art obfuscation schemes.
%
Meanwhile, it introduces no additional numerical error, preserving the computation correctness, and enables flexible trade-off between protection and performance.
In particular, \flob\ reduces the operator recovery rate of \btd, a representative reverse engineering analyzer, to 4.51\%, achieving a reduction of more than 30 percentage points relative to \nnshield, a state-of-the-art obfuscator for DNN binaries.

In summary, the paper makes the following contributions:
\begin{itemize}
	\item We propose \flob, an obfuscation scheme to protect floating-point computations in DNN binaries with MBA obfuscation.
	It reinterprets floating-point numbers in a higher-precision representation space, which enables bitwise transformations without violating numerical semantics.
	
	\item \flob\ performs computation in a lifted domain and restores results only at the end, preserving correctness without introducing additional numerical error.
	
	\item We have implemented a prototype of \flob\ and evaluated it on real-world DNN models.
	The results show that it improves resistance against state-of-the-art reverse engineering and deobfuscation techniques while preserving computational correctness and allowing flexible trade-off between protection and performance.
	
\end{itemize}

\section{Background and Motivation}

In this section, we first review the representation and arithmetic properties of floating-point numbers under the IEEE-754 standard, which form the basis of DNN computation, and introduce mixed boolean-arithmetic~(MBA) transformations, a class of equivalence-preserving techniques widely used for obfuscating integer computations.
We then discuss the limitations of directly applying MBA to floating-point computation, highlighting the inherent gap between numerical semantics and representation-level operations.

\subsection{IEEE-754 Floating-Point Numbers}\label{sec:mtv:fpn}

\begin{figure}[t]
	\centering
	\includegraphics[scale=.4]{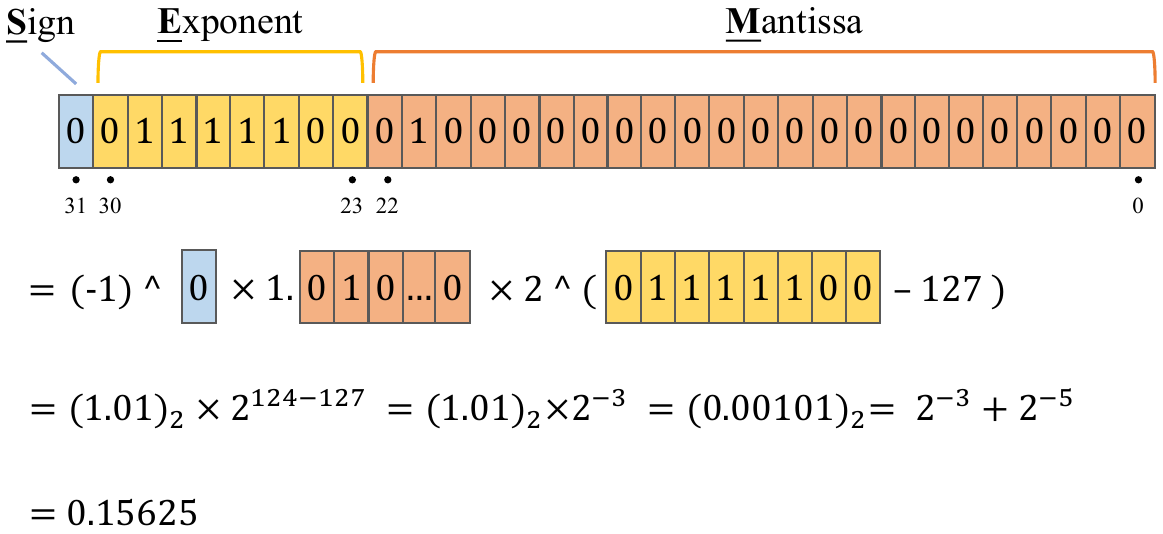}
	\caption{Example of IEEE-754 Floating-point Number in Single Precision}
	\label{fig:fpn-example}
\end{figure}

The IEEE-754 is a technical standard that defines how real-valued numbers are represented and computed in modern computing systems~\cite{ieee-754}.
It is designed to provide a finite-precision approximation of real numbers while ensuring well-defined and consistent numerical semantics across different platforms.
Such a representation is fundamental to DNN computation, where most operations are performed using floating-point formats.

Under IEEE-754, a floating-point number is encoded using three components:
a sign bit $S$ to indicate the sign of the number, an exponent $E$ to determine the scaling factor, and a mantissa $M$ to encode the significant bits of the number.
Then, a normalized single-precision~(FP32) number is represented as:
\begin{equation}
	FPN_{32} = (-1)^S \times (1.M) \times 2^{E-127},
	\label{eq:fpn32}
\end{equation}
where the constant $127$ denotes the exponent bias for FP32.
Figure~\ref{fig:fpn-example} shows an example of 32-bit floating-point number.
It can be decomposed into a sign bit $S=0$, an exponent field $E=01111100$, and a mantissa field of $M=010\ldots0$.
The corresponding value is:
$(-1)^0 \times (1.01)_2 \times 2^{124-127} = (1.01)_2 \times 2^{-3} = (0.00101)_2$, i.e.,~$0.15625$.

Notably, IEEE-754 represents real values using finite precision, and many decimal numbers, e.g.,~$0.1$ or $0.2$, cannot be represented exactly in binary form.
Arithmetic operations are therefore subject to rounding, introducing small numerical deviations depending on precision and evaluation order.
Despite this, the standard ensures consistent numerical semantics for floating-point computation.

\subsection{Bitwise Operations and MBA Obfuscation}
Bitwise operations manipulate data at the level of individual bits and form the basis of many low-level program transformations.
Common operators include \texttt{AND}, \texttt{OR}, and \texttt{XOR}, which operate at the bit level and follow basic Boolean logic.

Mixed boolean-arithmetic~(MBA)~\cite{mba-definition} transformations combine bitwise and arithmetic operations to construct expressions that are functionally equivalent but syntactically more complex.
For instance:

\begin{minipage}{0.45\linewidth}
	\begin{equation*}
		x - y =
		\begin{cases}
			x + (\neg y + 1) \\
			(x \oplus y) - 2 \cdot (\neg x \land y) \\
			(x \lor \neg y) - (x \land y)
		\end{cases}
	\end{equation*}
\end{minipage}
\hfill
\begin{minipage}{0.45\linewidth}
	\begin{equation*}
		x \oplus y =
		\begin{cases}
			(x \lor y) - (x \land y) \\
			(x \land \neg y) \lor (\neg x \land y) \\
			(x + y) - 2 \cdot (x \land y)
		\end{cases}
	\end{equation*}
\end{minipage}.
Such transformations preserve the original computation semantics while significantly altering the structure of the expression, making them widely used in code obfuscation, i.e,~MBA obfuscation.

From a perspective of software protection, 
obfuscation aims to increase the cost of reverse engineering rather than prevent program understanding entirely.
Although advanced deobfuscation techniques, such as solver- and synthesis-based simplification~\cite{mba-blast,xu2021boosting,promba}, are able to simplify such expressions in principle, MBA obfuscation typically incurs substantial computational and manual effort for these approaches, making large-scale and automated recovery less practical in real-world scenarios.


It is worth noting that the correctness of MBA transformations depends on a close alignment between representation-level operations and numerical semantics, which naturally holds for integer computation.
However, as discussed next,
this alignment does not directly extend to IEEE-754 floating-point computation, where the relationship between bit-level representation and numerical semantics is more intricate.

\subsection{MBA for Floating-Point Numbers}

Despite its effectiveness in software protection, MBA obfuscation cannot be extended to IEEE-754 floating-point numbers.
The limitation stems from the fundamental differences between bitwise computation and floating-point computation.
\begin{enumerate}[label=(\arabic*)]
	\item \textbf{Mismatched Semantics:}
	In IEEE-754, the bit pattern of a floating-point number does not directly correspond to its numerical value.
	As shown in Figure~\ref{fig:fpn-example}, a number is composed of structured fields, including Sign, Exponent, and Mantissa, and its value is determined through exponent scaling and normalization.
	Consequently, bitwise operations on the fields lead to unpredictable changes in the underlying value, breaking the numerical semantics required by correct obfuscation.
	
	\item \textbf{Rounding Errors:}
	Floating-point arithmetic in IEEE-754 is inherently approximate due to finite precision and rounding.
	As a result, the numerical outcomes vary with evaluation order and precision.
	This imprecision breaks the deterministic and semantics-preserving equivalence across different expressions, which is required by MBA obfuscation.	
\end{enumerate}

To address the above limitations,
\flob\ lifts floating-point values into a higher-precision representation space, where computations are performed while preserving numerical semantics, and only the final results are projected back to the target precision.
Therefore, it not only \textbf{aligns representation-level operations with numerical semantics}, but also \textbf{maintains sufficient precision throughout the computation}, eliminating intermediate rounding effects and preserving deterministic equivalence, thereby enabling MBA obfuscation for floating-point numbers.

\section{Design}

In this paper, 
\flob\ is designed to be integrated into the model compilation pipeline to protect sensitive computation before deployment.
It is intended for use by model developers, applying obfuscation to IEEE-754 floating-point computations at compilation time.
The resulting binary is then distributed for inference, serving as the deployment artifact without requiring any modification at runtime.
Notably, \flob\ maintains separate representations for values that would otherwise be indistinguishable under IEEE-754 rounding at the native precision,
and special cases, e.g.,~signed zeros and \texttt{NaN}s, are handled explicitly to preserve well-defined semantics.

\begin{figure*}[t]
	\includegraphics[scale=.7]{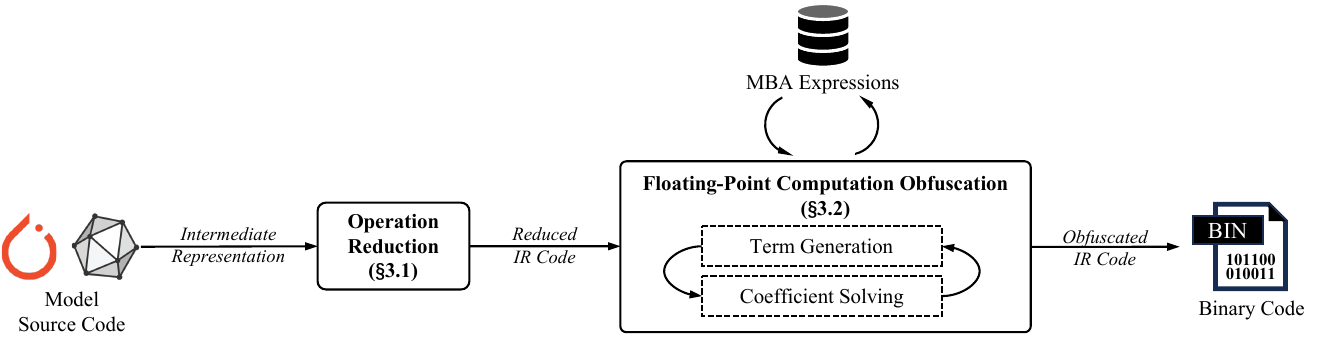}
	\caption{System Overview of \flob}
	\label{fig:sys-overview}
\end{figure*}

Figure~\ref{fig:sys-overview} illustrates the overall workflow of \flob.
Starting from model source code, \flob\ first lowers the computations into intermediate representation~(IR), and then decomposes original complex operations into reduced IR via \textbf{Operation Reduction~(\S\ref{sec:dsgn:reduce})}.
It subsequently replaces selected operations with semantically equivalent MBA expressions to achieve \textbf{Floating-Point Computation Obfuscation~(\S\ref{sec:dsgn:obf})}.
To this end, \flob\ maintains a repository of pre-constructed expressions and reuses them whenever they satisfy the transformation requirements.
Otherwise, it generates new expressions through \textit{Term Generation} and \textit{Coefficient Solving}, and incorporates them into the repository for future reuse.
Finally, the obfuscated IR code is compiled into binary code for deployment.

\subsection{Operation Reduction}\label{sec:dsgn:reduce}

\begin{algorithm}[t]
	\DontPrintSemicolon
	\caption{Reduced Floating-Point Multiplication}
	\label{alg:mul-reduce}
	
	\KwIn{The Left-Hand Side Operand $\mathit{P}$}
	\KwIn{The Right-Hand Side Operand $\mathit{Q}$}
	\KwOut{The Result of the Multiplication $\mathit{R}$}
	
	\Al{ReducedMultiply ($\mathit{P}$, $\mathit{Q}$)}{
		$s_p, e_p, m_p \leftarrow$ \texttt{DecomposeWithHiddenBit} ($\mathit{P}$) \label{alg:mul-reduce:opnd-dec-0} \\
		$s_q, e_q, m_q \leftarrow$ \texttt{DecomposeWithHiddenBit} ($\mathit{Q}$) \label{alg:mul-reduce:opnd-dec-1} \\
		$s_r \leftarrow s_p \oplus s_q$,
		$e_r \leftarrow e_p + e_q$,
		$m_r \leftarrow 0$ \label{alg:mul-reduce:se} \\
		\While{$m_q \ne 0$}{
			\lIf{$(m_q \land 1) = 1$}{
				$m_r \leftarrow m_r + m_p$ \label{alg:mul-reduce:acc}
			}
			$m_p \leftarrow m_p \ll 1$ \label{alg:mul-reduce:mp-shift} \\
			$m_q \leftarrow m_q \gg 1$ \label{alg:mul-reduce:mq-shift} \\
		}
		$R \leftarrow$ \texttt{ComposeWithHiddenBit} ($s_r$, $e_r$, $m_r$) \label{alg:mul-reduce:res} \\
		\KwRet{$\mathit{R}$}
	}
\end{algorithm}

After lowering model code into IR, floating-point computations are expressed as a mixture of operations with different algebraic properties.
In particular, many of them are inherently non-linear and cannot be directly transformed using MBA-based techniques, which rely on compositions of bitwise and linear arithmetic operations.
Therefore, to enable MBA obfuscation, \flob\ reduces non-linear operations, i.e.,~multiplication and division between variables, into sequences of linear ones.

Algorithm~\ref{alg:mul-reduce} presents the process to carry out reduced floating-point multiplication.
It first decomposes the two operands $P$ and $Q$ into their sign, exponent, and mantissa components~(\S\ref{sec:mtv:fpn}), where the mantissas are recovered together with the hidden bit~(\mbox{Line~\ref{alg:mul-reduce:opnd-dec-0}-\ref{alg:mul-reduce:opnd-dec-1}}).
The sign and exponent of the result are then derived from those of the two operands~(Line~\ref{alg:mul-reduce:se}).
The mantissa multiplication is reduced using the shift-and-add procedure~\cite{revol2000accelerated}.
The multiplier mantissa $m_q$ is scanned bit by bit, and whenever its least significant bit is $1$, the current multiplicand mantissa $m_p$ is accumulated into the running result $m_r$~(Line~\ref{alg:mul-reduce:acc}).
After each iteration, $m_p$ and $m_q$ are shifted separately, until all bits of $m_q$ have been processed~(\mbox{Line~\ref{alg:mul-reduce:mp-shift}-\ref{alg:mul-reduce:mq-shift}}).
Finally, the resulting floating-point value is reconstructed from the computed sign, exponent, and mantissa~(Line~\ref{alg:mul-reduce:res}).

The reduction of floating-point division follows the same principle.
The sign is still determined by the two operand signs, and the exponent is obtained by exponent subtraction, while the mantissa of the result is computed through the restoring-division method~\cite{robertson2009new}.
In this way, \flob\ generates semantically equivalent IR code to replace the original floating-point multiplication and division operations in the target IR, providing a uniform basis for later obfuscation.

\subsection{Floating-Point Computation Obfuscation}\label{sec:dsgn:obf}
This section presents the core design of \flob\ for obfuscating floating-point computations. 
It performs transformations in a lifted representation domain, where floating-point values are manipulated through their binary expansions while preserving numerical semantics~(\S\ref{sec:dsgn:obf:bitwise}).
Based on the representation, \flob\ constructs MBA expressions using bitwise and linear arithmetic operations, then replaces target operations with their obfuscated counterparts.
Specifically, the construction consists of two steps:
\begin{enumerate*}[label=\roman*)]
	\item generating candidate terms which define the structure of the expression~(\S\ref{sec:dsgn:obf:term}); and
	
	\item determining the coefficients of the terms to ensure semantic equivalence~(\S\ref{sec:dsgn:obf:solving}).
\end{enumerate*}
Finally, a running example is provided to o illustrate the complete obfuscation process~(\S\ref{sec:dsgn:obf:example}).

\subsubsection{Floating-Point Representation in \flob}\label{sec:dsgn:obf:bitwise}
Since the native IEEE-754 encoding is unsuitable for direct manipulation with bitwise operations,
\flob\ lifts the floating-point values and represents them by the binary expansion.
Specifically, a value $x$ is expressed as a finite sum of scaled binary digits:
\begin{equation}
	x = \pm (b_mb_{m-1}\cdots b_0 \mathpunct{.} b_{-1}b_{-2}\cdots b_l) = s\cdot\sum_{i=l}^m b_i\cdot 2^i,
	\label{eq:bitstring}
\end{equation}
where $b_i\in\{0, 1\}$, $s\in\{+1,-1\}$ denotes the sign, and $m$ and $l$ are determined by the precision of the original value.
For special numbers, $\pm0$ are treated as a single value $0$, while $\pm$\texttt{Inf} and \texttt{NaN} are preserved and processed separately when encountered.
Based on the representation,
\flob\ then handles \textit{arithmetic} and \textit{bitwise} operations as follows.

%
\textbf{\emph{Arithmetic Operations.}}
\flob\ considers only addition and scalar multiplication, which form the basis of MBA expressions. Subtraction is treated as addition with negative operand.
It follows the general procedure of floating-point arithmetic, where alignment precedes the computation, while operating entirely on the binary expansion without rounding.
For addition, operands are first aligned to share the same index range, i.e.,~that of $i$ in Equation~\ref{eq:bitstring}.
The result is then obtained by combining corresponding digits, with carries propagated across positions when necessary.
Scalar multiplication is realized as a combination of shifts and additions over the binary expansion.
For example, $$11 \cdot 0.11 = (0.11 \ll 1) + 0.11 = 1.1 + 0.11 =  10.01.$$
%

\textbf{\emph{Bitwise Operations.}}
Bitwise operations are not defined for floating-point number under IEEE-754 standard.
Nevertheless, they can be meaningfully interpreted at the representation level by leveraging the binary expansion in Equation~\ref{eq:bitstring}.
Specifically, given a floating-point value, \flob\ reinterprets its fractional expansion under a fixed precision as an integer representation, performs the corresponding bitwise operation on the integer domain, and then projects the result back to the floating-point space with the same scaling factor.
In this sense, the computation is purely representation-driven rather than mathematically defined over real numbers.
For example, consider the operation $3.75 \oplus 1.5$, i.e.,~$11.11 \oplus 1.1$ in binary expansion form.
Under a fixed 5-bit representation with 2 fractional bits, where the sign bit is included, the operands are scaled to integers as $01111$ and $00110$, respectively. 
The computation procedure is as follows:
\begin{equation*}
	11.11 \oplus 1.1 = \frac{01111 \oplus 00110}{2^2} = \frac{01001}{2^2} = 10.01.
\end{equation*}
\textbf{As long as a sufficient precision is provided, the above computation does not introduce additional numerical error}, since all operations are performed on exact integer representations and the scaling is deterministic.
In practice, the bit-width is chosen to be large enough to cover the representable range of the target floating-point format.
For instance, 129 bits are sufficient for single-precision floating-point numbers, if only the maximum magnitude is considered. 
However, if all IEEE-754 single-precision values, including subnormals, are to be represented exactly under a unified fixed scaling, the representation must also preserve the minimum precision of $2^{-149}$.
In that case, 278 bits are required in total, including one sign bit.

\begin{figure*}[t]
	\centering
	\begin{minipage}[t]{.45\textwidth}
		\centering
		\vspace{0pt}
		\begin{subfigure}[t]{\linewidth}
			\centering
			\includegraphics[scale=.5]{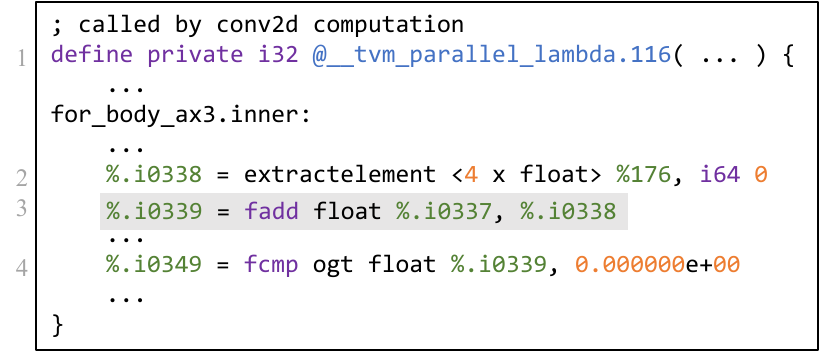}
			\caption{LLVM IR of the Floating-Point Addition}
			\label{fig:design:example:src-ir}
		\end{subfigure}

		\begin{subfigure}[t]{\linewidth}
			\centering
			\begin{subfigure}[t]{0.48\linewidth}
				\centering
				{\small
				\[
				((aux \land opnd_0) \oplus opnd_1) \lor opnd_0
				\]
				}
			\end{subfigure}
			\hfil
			\begin{subfigure}[t]{0.48\linewidth}
				\centering
				{\small
				\[
				(\neg(opnd_1 \oplus opnd_0)) \land opnd_1
				\]
				}
			\end{subfigure}
			\caption{Generated Candidate Terms}
			\label{fig:design:example:terms}
		\end{subfigure}
		
		\vspace{.8em}
		
		\begin{subfigure}[t]{\linewidth}
			\centering
			\fontsize{8}{8.8}\selectfont
			\renewcommand{\arraystretch}{0.97}
			\setlength{\tabcolsep}{10pt}
			\begin{tabular}{cc|c|cc}
				\toprule
				$\mathbf{opnd_0}$ & $\mathbf{opnd_1}$ & $\mathbf{aux}$ & $\mathbf{term_0}$ & $\mathbf{term_1}$ \\
				\midrule
				0 & 0 & 0 & 0 & 0 \\
				0 & 0 & 1 & 0 & 0 \\
				0 & 1 & 0 & 1 & 0 \\
				0 & 1 & 1 & 1 & 0 \\
				1 & 0 & 0 & 1 & 0 \\
				1 & 0 & 1 & 1 & 0 \\
				1 & 1 & 0 & 1 & 1 \\
				1 & 1 & 1 & 1 & 1 \\
				\bottomrule
			\end{tabular}
			\caption{Truth Table of Candidate Terms}
			\label{fig:design:example:truth-table}
		\end{subfigure}

	\end{minipage}
	\hfil
	\begin{minipage}[t]{.53\textwidth}
		\centering
		\vspace{0pt}
		\begin{subfigure}[t]{\linewidth}
			\includegraphics[scale=.5]{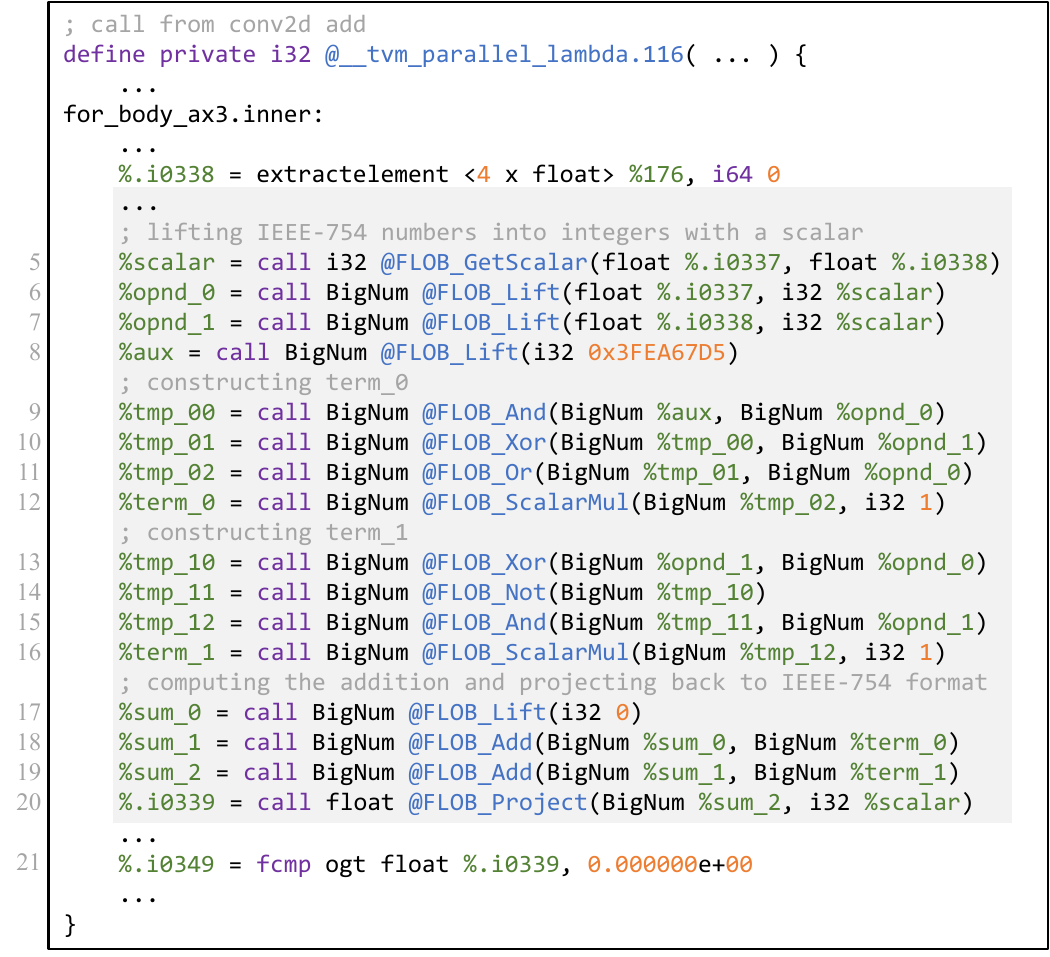}
			\caption{Obfuscated IR of the Floating-Point Addition}
			\label{fig:design:example:obf-ir}
		\end{subfigure}
	\end{minipage}
	\caption{Example of \flob\ applied to obfuscate a floating-point addition used during a Conv2d computation. The candidate terms are generated under the configuration of 2 terms, 1 auxiliary variable, and 3 operators per term.}
	\label{fig:design:example}
\end{figure*}

\subsubsection{Term Generation}\label{sec:dsgn:obf:term}

\begin{algorithm}[t]
	\DontPrintSemicolon
	\caption{Term Generation via Sampling}
	\label{alg:term-gen}
	
	\KwIn{Target Statement $\mathit{S}$}
	\KwIn{Required Number of Operators in a Term $\mathit{N_o}$}
	\KwIn{Required Number of Auxiliary Variables in a Term $\mathit{N_a}$}
	\KwIn{Set of Available Operators $\mathit{O}$}
	\KwOut{A Candidate Term $\mathit{T_c}$}
	
	\Al{GenerateOneTerm ($\mathit{S}$, $\mathit{N_o}$, $\mathit{N_a}$)}{
		$a \leftarrow$ \texttt{GenerateAuxiliaryVariable} ($\mathit{N_a}$) \label{alg:term-gen:av-gen} \\
		$T_c \leftarrow$ \texttt{SampleVariable} ($\mathit{S}$, $\mathit{a}$) \label{alg:term-gen:var-sel} \\
		$i \leftarrow 0$ \\
		\While{$i < \mathit{N_o}$\label{alg:term-gen:loop-cond}}{
			$o \leftarrow$ \texttt{SampleOperator} ($\mathit{O}$) \label{alg:term-gen:op-sel} \\
			$T_c \leftarrow$ \texttt{UpdateTerm} ($T_c$, $o$) \label{alg:term-gen:update-1} \\
			\If{$o$ is a binary operator\label{alg:term-gen:if-cond}}{
				$v \leftarrow$ \texttt{SampleVariable} ($\mathit{S}$, $\mathit{a}$) \\
				$T_c \leftarrow$ \texttt{UpdateTerm} ($T_c$, $v$) \label{alg:term-gen:update-2}
			}
			$i \leftarrow i + 1$ \\
		}
		\KwRet{$T_c$}\label{alg:term-gen:ret}
		
	}
\end{algorithm}

\flob\ generates candidate terms that define the structural components of an MBA expression.
The generation process is guided by configurable obfuscation requirements, including \emph{the number of operators} within each term and \emph{the number of auxiliary variables} to introduce redundancy.
These constraints control the complexity and diversity of the generated terms, providing a flexible trade-off between complexity and overhead.

The process of the generation is described in Algorithm~\ref{alg:term-gen}.
\flob\ first generates a set of auxiliary variables of size $N_a$, which specifies the number of such variables in a term~(Line~~\ref{alg:term-gen:av-gen}).
It then initializes the term $T_c$ by randomly sampling a variable from the union of variables in the target IR statement $S$ and the auxiliary variable set~(Line~\ref{alg:term-gen:var-sel}).
Afterwards, \flob\ incrementally constructs the term through a sequence of operator applications.
In each iteration, an operator is randomly sampled from the available operator set $O$, and applied to the current term~(\mbox{Lines~\ref{alg:term-gen:op-sel}–\ref{alg:term-gen:update-1}}).
If the selected operator is binary, an additional variable is sampled from the same variable pool and incorporated into the term~(\mbox{Lines~\ref{alg:term-gen:if-cond}-\ref{alg:term-gen:update-2}}).
This process continues until the number of applied operators reaches the required bound $N_o$~(Line~\ref{alg:term-gen:loop-cond}).
At last, the candidate term $T_c$ is returned~(Line~\ref{alg:term-gen:ret}), which serves as a building block for subsequent MBA expression construction.

\textbf{Example.}
Figure~\ref{fig:design:example} presents an example of \flob\ to obfuscate an floating-point addition from a Conv2d computation, which appears in the helper function \texttt{\_\_tvm\_parallel\_lambda.116}~(\mbox{Line~3}, Figure~\ref{fig:design:example:src-ir}).
The candidate terms are generated under the configuration of two terms, one auxiliary variable~($N_a=1$), and three operators per term~($N_o=3$), as shown in Figure~\ref{fig:design:example:terms}, where the two operands are denoted as $opnd_0$ and $opnd_1$, and the auxiliary variable is $aux$.
Taking the first term as an example, \flob\ first samples $aux$ to initialize the term. 
It then iteratively samples operators from $\{\land, \lor, \lnot, \oplus\}$ to extend the term accordingly.
$\land$ is selected in the example, and since it is binary, $opnd_0$ is sampled and combined with the current term, yielding $(aux\land opnd_0)$.
The term is enclosed in parentheses to preserve the intended evaluation order.
The process is repeated until the three operators are introduced, producing the final term $((aux\land opnd_0)\oplus opnd_1)\lor opnd_0$.
The second term, $(\lnot(opnd_1 \oplus opnd_0))\land opnd_1$, is generated in the same manner.

\subsubsection{Coefficient Solving}\label{sec:dsgn:obf:solving}

\begin{algorithm}[t]
	\DontPrintSemicolon
	\caption{Coefficient Solving with Linear Systems}
	\label{alg:solving}
	
	\KwIn{Target Statement $\mathit{S}$}
	\KwIn{Required Number of Terms $\mathit{N_t}$}
	\KwIn{Required Number of Operators in a Term $\mathit{N_o}$}
	\KwIn{Required Number of Auxiliary Variables in a Term $\mathit{N_a}$}
	\KwIn{Set of Available Operators $\mathit{O}$}
	\KwOut{Set of Coefficients of Candidate Terms $\mathit{C}$}
	
	\Al{SolveCoefficient ($S$, $T$)}{
		\Repeat{$C$ has a solution\label{alg:solving:repeat-cond}}{
			$n \leftarrow 0$, $t_c \leftarrow \{\ \}$ \\
			\While{$n < N_t$\label{alg:solving:loop-cond}}{
				$t_c \leftarrow t_c\ +$ \texttt{GenerateOneTerm} ($S$, $N_o$, $N_a$) \\
				$n \leftarrow n + 1$ \label{alg:solving:cnt-update}
			}
			$m \leftarrow$ \texttt{ConstructOneBitTruthTable} ($S$, $t_c$) \label{alg:solving:truth-table} \\
			$C \leftarrow$ \texttt{SolveLinearSystem} ($m$, $C$) \label{alg:solving:solve}
		}
		\KwRet{$C$} \label{alg:solving:ret}
	}
\end{algorithm}

Given the candidate terms, \flob\ next determines their coefficients so that the resulting linear combination is semantically equivalent to the original target statement.
Supposing that the candidate terms are $t_0, \ldots, t_k$,
\flob\ seeks coefficients $c_0, \ldots, c_k$, such that $$\sum_{i=0}^k c_i\cdot t_i \equiv S,$$
where $S$ is the result of the original statement to be obfuscated.

The solving is carried out in the integer domain after floating-point values have been lifted and scaled~(\S\ref{sec:dsgn:obf:bitwise}).
\flob\ then solves the coefficients only over 1-bit operands, because, for linear MBA identifiers, equivalence on 1-bit variables also holds for integers of arbitrary bit width under the same expression form~\cite{mba-blast,mba-definition}.
Therefore, \flob\ enumerates the truth table of the generated terms and the target expression in 1-bit space, deriving the corresponding linear constraints, and solves for a valid coefficient vector.
The resulting coefficients are then used to instantiate the final MBA expression over the lifted integer representation.
Algorithm~\ref{alg:solving} solves the coefficients of candidate terms through an iterative generate-and-solve procedure.
It starts by repeatedly generating candidate terms until the required number of terms $N_t$ is obtained~(\mbox{Line~\ref{alg:solving:loop-cond}-\ref{alg:solving:cnt-update}}).
\flob\ then constructs the 1-bit truth table of both the target statement $S$ and the candidate terms~(Line~\ref{alg:solving:truth-table}), deriving a linear system whose unknowns are the coefficients of the candidate terms, and solves the system accordingly~(Line~\ref{alg:solving:solve}).
The process continues until a valid coefficient set is found~(Line~9), which is then returned as the solution for the final MBA expression~(Line~\ref{alg:solving:ret}).

\textbf{Example.}
Given the candidate terms~(Figure~\ref{fig:design:example:terms}),
\flob\ enumerates all possible values of $opnd_0$, $opnd_1$, and $aux$ in 1-bit space, thereby constructing the truth table of $term_0$ and $term_1$, as shown in Figure~\ref{fig:design:example:truth-table}.
It then builds the linear equation system as follows, where the left-hand side corresponds to the candidate MBA expression $c_0\cdot term_0 + c_1\cdot term_1$, and the right-hand side is derived from the target statement $opnd_0 + opnd_1$.
The system is solved to yield the coefficient vector $[1, 1]^{\top}$.
\begin{equation*}
	\begin{bNiceArray}{cc}[first-row]
		\mathbf{term_0} & \mathbf{term_1} \\
		0 & 0 \\
		0 & 0 \\
		1 & 0 \\
		1 & 0 \\
		1 & 0 \\
		1 & 0 \\
		1 & 1 \\
		1 & 1
	\end{bNiceArray}
	\cdot
	\begin{bmatrix}
		c_0 \\
		c_1
	\end{bmatrix}
	=
	\begin{bNiceArray}{cc}[first-row]
		\mathbf{opnd_0} & \mathbf{opnd_1} \\
		0 & 0 \\
		0 & 0 \\
		0 & 1 \\
		0 & 1 \\
		1 & 0 \\
		1 & 0 \\
		1 & 1 \\
		1 & 1
	\end{bNiceArray}
	\cdot
	\begin{bmatrix}
		1 \\
		1
	\end{bmatrix}
\end{equation*}

\subsubsection{Running Example}\label{sec:dsgn:obf:example}
Figure~\ref{fig:design:example} summarizes the end-to-end obfuscation process of \flob\ using a floating-point addition extracted from a Conv2d computation.
Given the addition statement in Figure~\ref{fig:design:example:src-ir}, it generates candidate terms 
through randomized sampling~(Figure~\ref{fig:design:example:terms}), and constructs the 1-bit truth table for coefficient solving~(Figure~\ref{fig:design:example:truth-table}), obtaining the solution vector $[1, 1]^{\top}$ for the two terms.
Figure~\ref{fig:design:example:obf-ir} shows the resulting obfuscated IR of the target operation.
The original floating-point operands, together with the auxiliary variable, are first lifted into the integer domain with a scaling factor~(\mbox{Line~5-8}), where \texttt{BigNum} is the data structure to represent the lifted integers.
Then, the two terms are instantiated through the corresponding bitwise and linear arithmetic operations, and combined with their solved coefficients to form the MBA expression~(\mbox{Line~9-16}).
Finally, the transformed computation is evaluated and projected back to the IEEE-754 format to replace the original floating-point addition~(\mbox{Line~17-20}).

\section{Implementation}


The prototype of \flob\ supports the obfuscation of double-precision floating-point computation in DNN models on the Linux platform.
It adopts \texttt{TVM}~\cite{tvm} as the front-end to transform model files, e.g.,~\texttt{ONNX}, into LLVM IR.
\flob\ then performs obfuscation on the IR code, which is developed in C++ with about 7,400 lines of code.
Finally, it leverages the \texttt{LLVM} framework~\cite{llvm} as the back-end to compile the obfuscated IR into binaries.

To handle double-precision numbers,
\flob\ uses a lifted representation with fixed bounds $l$ and $m$, as shown in Equation~\ref{eq:bitstring}, allocating sufficient bits for both the integer and fractional parts.
Additionally,
\flob\ provides three configurable parameters, i.e.,~the number of terms, the number of operators per term, and the number of auxiliary variables, to control the structure and complexity of generated MBA expressions, as described in Algorithm~\ref{alg:solving}~(\S\ref{sec:dsgn:obf:solving}).

\section{Evaluation}

In this section, we conduct empirical experiments to evaluate the effectiveness and capacity of \flob\ with the following research questions~(RQs):
\begin{itemize}
	\item \textbf{RQ1: Correctness.}
	Does \flob\ preserve the correctness of the original floating-point computation without introducing additional numerical error?
	
	\item \textbf{RQ2: Effectiveness.}
	How effective is \flob\ in resisting existing reverse engineering and deobfuscation techniques?
	
	\item \textbf{RQ3: Performance and Overhead.}
	What is the practical cost of \flob\ in terms of obfuscation cost and runtime overhead?
\end{itemize}

\subsection{Evaluation Setup}
%
\flob\ uses $l=-1120$ and $m=1119$ in Equation~\ref{eq:bitstring}, so that both the integer part and the fractional part are allocated 1120 bits.
This range is sufficient to cover the full value range of IEEE-754 double-precision numbers, from the largest finite value on the order of $2^{1024}$ to the smallest positive subnormal value $2^{-1074}$, while preserving enough precision for the transformed computation.
%
In addition, unless otherwise specified, \flob\ uses the default configuration of $N_t=4$, $N_o=4$, and $N_a=2$ throughout the evaluation to generate MBA expressions, where the three parameters denote the number of terms, operators per term, and auxiliary variables respectively~(Algorithm~\ref{alg:solving} in \S\ref{sec:dsgn:obf:solving}).
This setting is used for all models to ensure consistency.

The evaluation is performed on the machine with AMD Ryzen Threadripper 3970X CPU @ 3.7GHz, 256GB memory, and Ubuntu 22.04.

\subsubsection{Model Set and Datasets}

\begin{table*}[t]
	\centering
	
	\caption{DNN Model Set and Test Input Datasets}
	\label{tab:dataset}
	\begin{tabular}{cccccccc}
		\toprule
		\multirow{2}{*}{Model} & \multirow{2}{*}{Type} & \multirow{2}{*}{\tabincell{c}{Test Input\\Dataset}} & \multirow{2}{*}{\tabincell{c}{No. of\\Operations}} & \multirow{2}{*}{FP Ratio} & \multirow{2}{*}{\tabincell{c}{Execution Time\\(ms)}} & \multirow{2}{*}{\tabincell{c}{Runtime Memory\\(GB)}} & \multirow{2}{*}{\tabincell{c}{Binary Size\\(MB)}} \\
		&&&&&&& \\ \midrule\midrule
		
		AlexNet & CV-CNN & ImageNet & 2,404 & 10.7\% & 44.510 & 1.484 & 239.116 \rule{0pt}{2.5ex} \\
		
		\rowcolor[HTML]{EFEFEF}
		VGG16 & CV-CNN & ImageNet & 4,095 & 16.2\% & 140.180 & 2.373 & 541.166 \rule{0pt}{2.5ex} \\

		ResNet18 & CV-CNN & ImageNet & 6,817 & 33.9\% & 15.620 & 0.985 & 46.859 \rule{0pt}{2.5ex} \\
		
		\rowcolor[HTML]{EFEFEF}
		MobileNet V2 & CV-CNN & ImageNet & 8,428 & 26.6\% & 23.430 & 0.928 & 16.928 \rule{0pt}{2.5ex} \\
		
		MNIST & CV-CNN & MNIST & 1,163 & 8.7\% & 0.260 & 0.860 & 1.830 \rule{0pt}{2.5ex} \\
		
		\rowcolor[HTML]{EFEFEF}
		RoBERTa & NLP-Transformer & SST-2 & 4,061 & 33.6\% & 1014.800 & 1.838 & 488.034 \rule{0pt}{2.5ex} \\
		
		\multirow{2}{*}{DistilBERT} & \multirow{2}{*}{NLP-Transformer} & \multirow{2}{*}{\tabincell{c}{Twitter US Airline\\Sentiment}} & \multirow{2}{*}{3,944} & \multirow{2}{*}{32.9\%} & \multirow{2}{*}{499.700} & \multirow{2}{*}{1.380} & \multirow{2}{*}{260.996} \rule{0pt}{2.5ex} \\
		&&&&&&& \rule{0pt}{2.5ex} \\
		\bottomrule
		
	\end{tabular}
\end{table*}

Table~\ref{tab:dataset} summarizes the evaluated model set and the corresponding test input datasets.
We use seven real-world DNN models, including five CV-CNN models and two NLP-Transformer models.
For each model, we use a task-relevant test input dataset as the input source for evaluating the obfuscated binary, so as to evaluate correctness and practical overhead under realistic inference workloads.

The selected models also exhibit a nontrivial proportion of floating-point operations, with FP ratios ranging from 8.7\% to 33.9\%.
The data is collected and measured on the corresponding LLVM IR, indicating that floating-point computation constitutes an important part of real-world DNN binaries, thereby justifying the need to protect floating-point semantics in practice.
The last three columns of the table report the pre-obfuscation execution time, runtime memory usage, and executable binary size of each model, which serve as the baseline for quantifying the overhead introduced by \flob\ after obfuscation.
In addition, the model set used in \nnshield~\cite{neuroshield} is adopted as well in the comparison experiments.

\subsubsection{Baseline Methods}
We adopt the state-of-the-art and representative obfuscation schemes, deobfuscation methods, and reverse engineering techniques as baselines, including:
\begin{itemize}
	\item \nnshield~\cite{neuroshield}, a state-of-the-art obfuscation framework to protect DNN binaries against reverse engineering.
	\item \xsmir~\cite{xsmir}, a state-of-the-art black-box deobfuscation tool for recovering the semantics of heavily obfuscated code through program synthesis.
	\item \gamba~\cite{gamba}, a representative MBA deobfuscation tool based on algebraic rewriting.
	\item \promba~\cite{promba}, a representative MBA deobfuscation tool that combines program synthesis and term rewriting.
	\item \btd~\cite{btd}, a state-of-the-art reverse engineering method to generally recover operator-level semantics from DNN binaries.
\end{itemize}

Additionally, we consider several other related methods which are not included in the final evaluation.
We further discuss the differences between \flob\ and these excluded methods in Section~\ref{sec:disc:compare}.
\loki~\cite{loki} is a virtual machine-based obfuscation framework strengthened with extensive MBA transformations, and \tigress~\cite{tigress} is a source-to-source obfuscator for C programs supporting a broad range of anti-analysis transformations, including MBA obfuscation.
Since both target traditional software protection over integer computations and control-flow structures, rather than floating-point computation obfuscation, we exclude them from direct comparison.

\texttt{NeuReduce}~\cite{feng-etal-2020-neureduce} is a neural-network-based MBA reducer for simplifying obfuscated MBA expressions.
\mbox{\texttt{MBA-Blast}}~\cite{mba-blast} simplifies MBA expressions by exploiting the equivalence between 1-bit and n-bit MBA transformations.
\mbox{\texttt{MBA-Flatten}} performs in-place MBA simplification by rewriting bitwise subexpressions into a unified form followed by arithmetic reduction.
\mbox{\texttt{MBA-Solver}}~\cite{xu2021boosting} simplifies MBA expressions by reducing the mixing degree of bitwise and arithmetic operations for easier solving.
\texttt{Syntia}~\cite{syntia} uses program synthesis to recover simplified semantics from obfuscated code, including MBA-obfuscated expressions.
As these methods are all representative traditional MBA deobfuscation approaches and their functionality is already covered by the selected baseline \gamba\ and \promba, they are not included separately in the final evaluation.

\mdobf~\cite{modelobfuscator} obfuscates model structure, parameters, and attributes to protect deployed TFLite models on mobile devices, and \dynamo~\cite{dynamo} extends this line of work with homomorphic encryption~\cite{yi2014homomorphic} against runtime instrumentation attacks. 
Since both are developed for on-device TFLite deployments rather than standalone executable binaries, we exclude them from the experimental baselines.

\dnd~\cite{dnd} is a DNN decompiler that recovers high-level DNN semantics from compiled binaries through symbolic execution and loop analysis, and \nnscope~\cite{neuroscope} is a dynamic-analysis-based framework for reversing DNN binaries on edge devices.
Since \btd\ already serves as a representative baseline for reversing DNN binaries as \dnd, we omit \dnd\ in the evaluation.
In addition, the released artifacts of \nnscope\ lack sufficient usage and configuration guidance, and its target programs depend on the \texttt{Glow}~\cite{rotem2018glow}, whose official repository has been archived and is now read-only~\cite{glow-github}, indicating that it is no longer actively maintained.
Therefore, we also exclude \nnscope\ from the evaluation.

\subsection{RQ1: Correctness}

In this section, \flob\ is evaluated to determine whether it preserves computation correctness without introducing additional numerical error.
The correctness is first examined in transforming basic floating-point operations~(\S\ref{sec:eva:rq1:basic}), and further in preserving the functional correctness of full DNN models after obfuscation~(\S\ref{sec:eva:rq1:model}).


In both cases,
the three obfuscation parameters are varied over $[2,6]$, $[2, 6]$, $[0, 4]$ separately, which are centered around the default configuration $(N_t, N_o, N_a) = (4, 4, 2)$, yielding overall $125$ obfuscation configurations.
Then, under each configuration, the target is obfuscated and tested with $100$ independently sampled inputs, resulting in $12,500$ tests in total.
For each test, the output of the obfuscated target is compared with that of the original one, and it is counted as correct only when the output difference is zero.

\begin{figure}[t]
	\centering
	\begin{subfigure}[t]{0.43\linewidth}
		\centering
		\hspace{-6mm}
		\includegraphics[scale=.44]{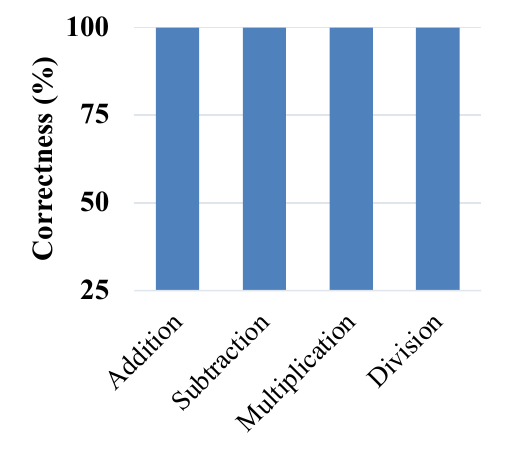}
		\caption{Operation}
		\label{fig:rq1:operation}
	\end{subfigure}
	\hfill
	\begin{subfigure}[t]{0.56\linewidth}
		\centering
		\includegraphics[scale=.44]{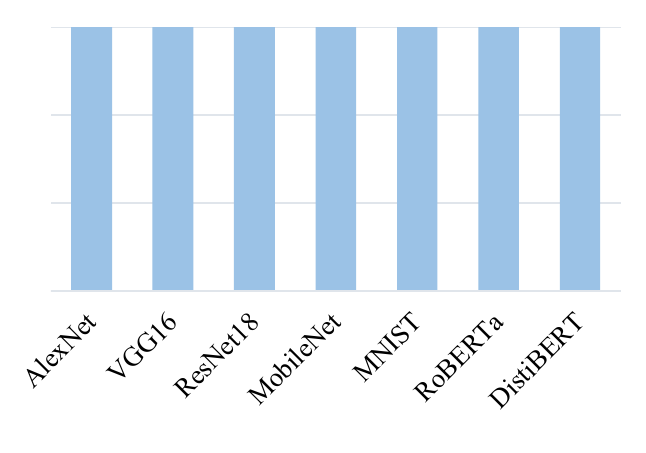}
		\caption{Model}
		\label{fig:rq1:model}
	\end{subfigure}
	\caption{Correctness Evaluation of \flob}
	\label{fig:rq1}
\end{figure}

\subsubsection{Floating-Point Operation Obfuscation}\label{sec:eva:rq1:basic}
The results are shown in Figure~\ref{fig:rq1:operation}.
Across all evaluated operations, a correctness rate of 100\% is consistently achieved.
This indicates that no additional numerical error is introduced by \flob\ during obfuscation.
The reason is that the floating-point values in MBA expressions are represented with higher precision~(\S\ref{sec:dsgn:obf:bitwise}), so that the transformations can be applied without perturbing the original semantics.
%
Additionally, the results of multiplication and division further show that the operation reduction is also semantics-preserving~(\S\ref{sec:dsgn:reduce}).

\subsubsection{DNN Model Obfuscation}\label{sec:eva:rq1:model}
As presented in Figure~\ref{fig:rq1:model},
\flob\ achieves a correctness rate of 100\% for all evaluated DNN models as well.
In this experiment, correctness is determined by comparing the output vectors of the original and obfuscated binaries.
Hence, every element in the output vector remains numerically identical after obfuscation, indicating that \flob\ preserves the end-to-end inference behavior of the original models without introducing additional numerical error.
Together with the results of operation obfuscation, it is further demonstrated that the correctness of \flob\ is maintained not only for individual floating-point operations but also for full DNN execution.

\begin{tcolorbox}[top=2.5pt, bottom=2.5pt, left=2.5pt, right=2.5pt]
	\textbf{Answer to RQ1}:
	\flob\ introduces no additional numerical error to original computation.
	Exact semantics is preserved for both individual floating-point operations and full DNN model inference, demonstrating the correctness of \flob\ at both the operator and model levels.
\end{tcolorbox}

\subsection{RQ2: Effectiveness}

In this section, the effectiveness of \flob\ is evaluated from the perspective of resistance to deobfuscation and reverse engineering.
Since it provides protection at both the floating-point expression level and the DNN binary level, two types of analysis are considered:
\begin{enumerate*}[label=\roman*)]
	\item the robustness of obfuscated floating-point expressions against MBA deobfuscation tools~(\S\ref{sec:eva:rq2:dobf}), and
	\item the resilience of obfuscated DNN binaries against reverse engineering attacks~(\S\ref{sec:eva:rq2:re}).
\end{enumerate*}

\subsubsection{MBA Deobfuscation}\label{sec:eva:rq2:dobf}

\begin{figure*}[t]
	\centering
	\begin{subfigure}[t]{.32\textwidth}
		\centering
		\hspace{-8mm}
		\includegraphics[scale=.45]{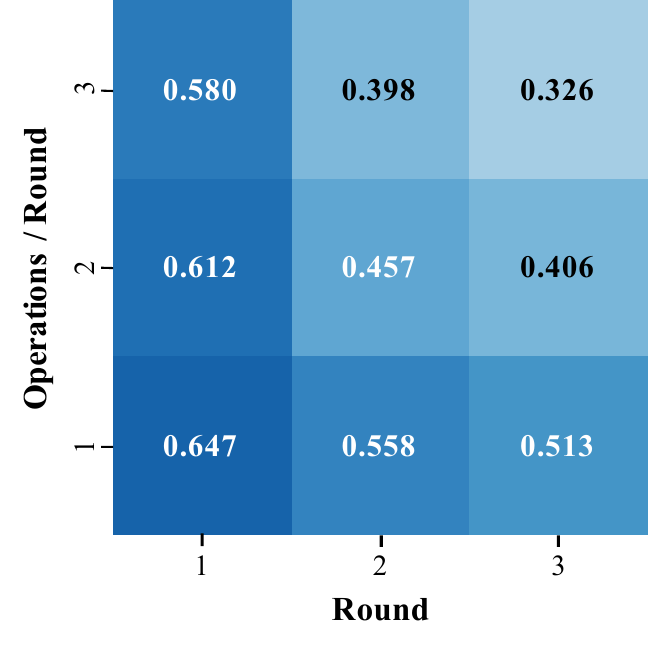}
		\caption{\gamba}
		\label{fig:rq2:gamba}
	\end{subfigure}
	\hfill
	\begin{subfigure}[t]{.32\textwidth}
		\centering
		\hspace{-8mm}
		\includegraphics[scale=.45]{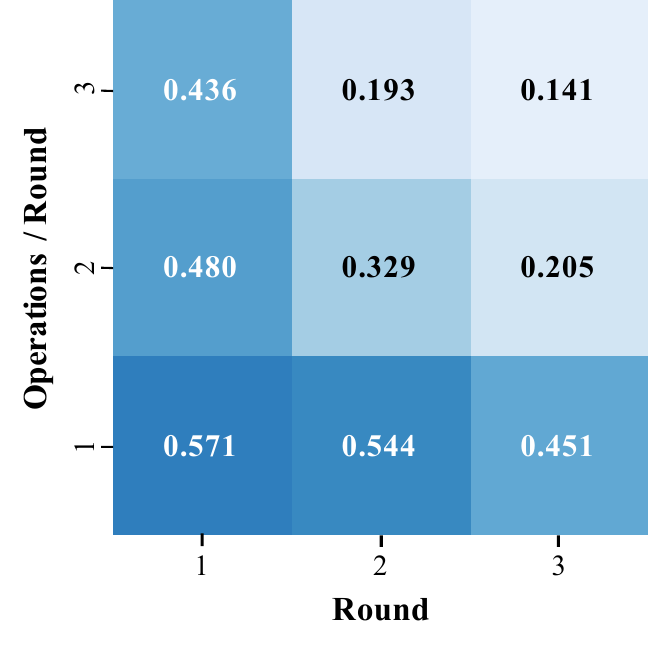}
		\caption{\promba}
		\label{fig:rq2:promba}
	\end{subfigure}
	\hfill
	\begin{subfigure}[t]{.32\textwidth}
		\centering
		\hspace{-8mm}
		\includegraphics[scale=.45]{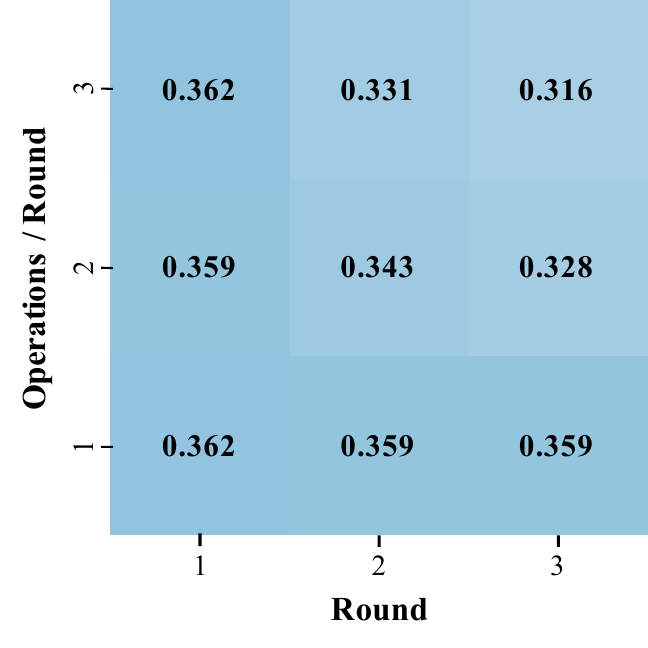}
		\caption{\xsmir}
		\label{fig:rq2:xsmir}
	\end{subfigure}
	\caption{Success rates of deobfuscation tools under different \flob\ configurations. Lower values indicate stronger protection. The x-axis denotes the round of obfuscation, and the y-axis shows the number of operations obfuscated in each round.}
	\label{fig:rq2}
\end{figure*}

Existing MBA deobfuscation methods are primarily designed for integer expressions, such that they cannot be directly applied to analyze full DNN binaries obfuscated by \flob.
Therefore, in this section, the evaluation is performed at the expression level, by constructing a dedicated test set of obfuscated expressions extracted from the evaluated DNN models.
Specifically, 200 target expressions are randomly sampled from the model set, as presented in Table~\ref{tab:dataset}.
To ensure that the extracted expressions reflect the characteristics of real DNN computation, the sampling ratio is aligned with the proportion of integer and floating-point operations in the corresponding models. Each sampled expression is then obfuscated by \flob\ and submitted to the evaluated MBA deobfuscation tools for analysis.

A deobfuscation attempt is regarded as successful only when the returned expression is semantically equivalent to the original one.
Otherwise, it is counted as a failure, including not only incorrect simplification results, but also cases where no valid result is returned, such as crashes, runtime errors, or timeouts.
For fairness, each deobfuscation task is assigned a timeout according to the default setting of the corresponding deobfuscator,
namely 3,600 seconds for \gamba~\cite{gamba}, 4,000 seconds for \promba~\cite{promba}, and 60 seconds for \xsmir~\cite{xsmir}.
The success rates of deobfuscation are shown in Figure~\ref{fig:rq2}, where round denotes the number of obfuscation iterations applied by \flob, and the other axis represents the number of operations obfuscated in each round.

Overall, the deobfuscation success rates of all three tools decrease as the \flob\ obfuscation becomes stronger, i.e., as more rounds are applied and more operations are obfuscated in each round.
This trend is consistently observed for the three deobfuscators, indicating that \flob\ effectively increases the difficulty of automated MBA deobfuscation under progressively stronger protection settings.
Even in the mildest configuration of \flob, the success rates are only 64.7\% for \gamba, 57.1\% for \promba, and 36.2\% for \xsmir, which are lower than the average accuracies reported in their original papers, namely 99.7\%, 84.4\%, and 76.3\%, respectively.

The lower success rates of all three tools can be attributed to the difference between their original design scope and the scenarios considered in this paper.
Existing MBA deobfuscators are primarily designed for integer-oriented expressions, whereas the expressions produced by \flob\ are derived from real-world DNN models with floating-point computations.
As a result, all three tools exhibit lower success rates when analyzing code obfuscated by \flob.
\xsmir\ remains relatively stable across different \flob\ configurations.
This is consistent with its black-box design, which attempts to recover expressions from input-output behaviors rather than directly analyzing expression structures.
Its performance is therefore less sensitive to changes in the obfuscation configuration of \flob.
Meanwhile, the expressions extracted from real-world models are complex and often involve large integers.
This further increases the difficulty of synthesis-based recovery, thereby reducing the overall success rate of \xsmir.

In contrast, \gamba\ and \promba\ are white-box methods that directly analyze expressions.
Their results therefore vary more clearly with different configurations of \flob.
Besides, 
\promba\ is more sensitive to analysis time, because solver-based reasoning is involved in its simplification process.
When the target expressions become more complex, it often cannot complete the analysis within the given time, leading to a lower observed success rate.
\gamba\ is less affected by this issue and thus performs better under the same time budget.
This observation is also consistent with the findings reported in the \xsmir\ paper~\cite{xsmir}.
\promba\ can outperform \gamba\ when sufficient time is available, whereas its advantage diminishes when the time budget is limited.

\subsubsection{DNN Reverse Engineering}\label{sec:eva:rq2:re}

\begin{figure}[t]
	\centering
	\includegraphics[scale=.59]{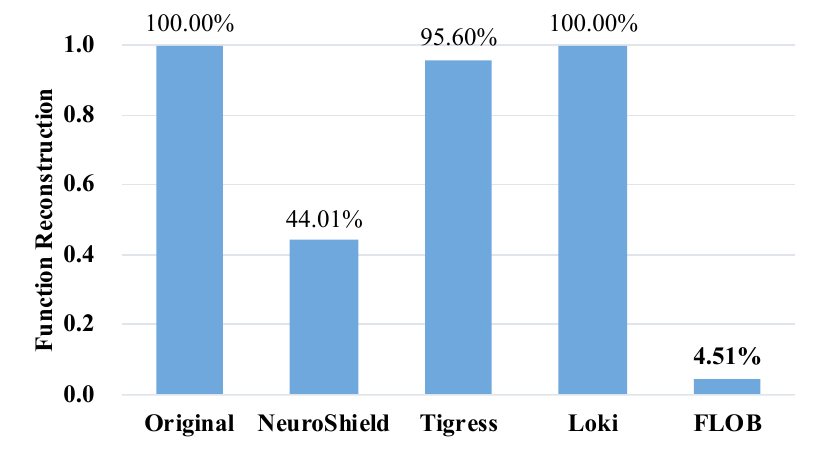}
	\caption{Function reconstruction rates by \btd. Lower values indicate stronger protection.}
	\label{fig:rq2:btd}
\end{figure}

Figure~\ref{fig:rq2:btd} presents the function reconstruction rates of \btd\ on original and obfuscated DNN binaries.
\nnshield\ and \btd\ are the state-of-the-art obfuscation scheme and reverse engineering technique for DNN binaries.
Therefore, we follow the evaluation configuration of \nnshield\ and reuse its model set as well for the comparison.
Specifically, the target models are obfuscated by \flob\ and then analyzed by \btd, and the ratio of successfully reconstructed functions is measured as the metric.
As a reference, Original shows the result on models without obfuscation, where \btd\ achieves a function reconstruction rate of 100.0\%.
The results of \nnshield, \tigress, and \loki\ are taken from the paper of \nnshield~\cite{neuroshield}.
In particular, the value reported for \nnshield\ corresponds to the average reconstruction rate over both CV and NLP models.
Under the same evaluation setting, \flob\ reduces the function reconstruction rate to only 4.51\%, which is substantially lower than those of other work.

These results indicate that \flob\ provides stronger protection against DNN binary reverse engineering than the compared obfuscation methods.
Unlike existing methods that mainly obfuscate code structure or conventional integer-oriented logic, \flob\ directly protects floating-point computations that are central to DNN execution.
As a result, the operator semantics and function behaviors required by \btd\ for reconstruction become much harder to recover.

\begin{tcolorbox}[top=2.5pt, bottom=2.5pt, left=2.5pt, right=2.5pt]
	\textbf{Answer to RQ2}:
	\flob\ effectively resists both MBA deobfuscation and DNN reverse engineering.
	It consistently reduces the success rates of state-of-the-art analysis tools, demonstrating strong protection at both the floating-point expression level and the DNN binary level.
\end{tcolorbox}

\subsection{RQ3: Performance and Overhead}
In this section, the practical performance and overhead of \flob\ are evaluated.
The analysis covers both the processing time incurred during the obfuscation~(\S\ref{sec:rq3:performance}) and the overhead introduced to the protected DNN binaries~(\S\ref{sec:rq3:overhead}).

\begin{figure*}[t]
	\centering
	\begin{subfigure}[t]{.32\textwidth}
		\centering
		\includegraphics[scale=.45]{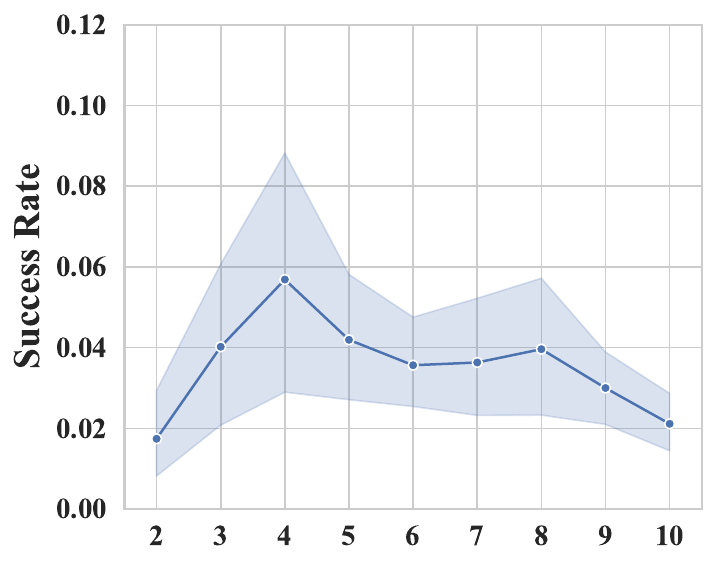}
		\caption{Term Number}
		\label{fig:rq3:term}
	\end{subfigure}
	\hfill
	\begin{subfigure}[t]{.32\textwidth}
		\centering
		\includegraphics[scale=.45]{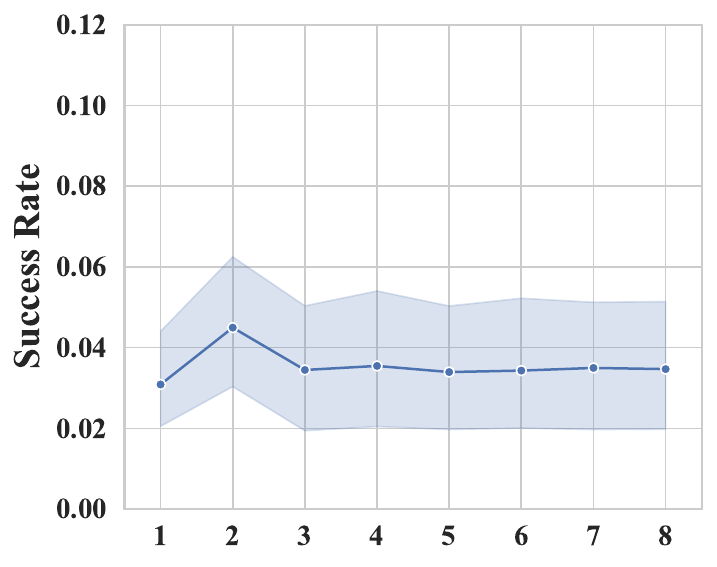}
		\caption{Operator Number per Term}
		\label{fig:rq3:operator}
	\end{subfigure}
	\hfill
	\begin{subfigure}[t]{.32\textwidth}
		\centering
		\includegraphics[scale=.45]{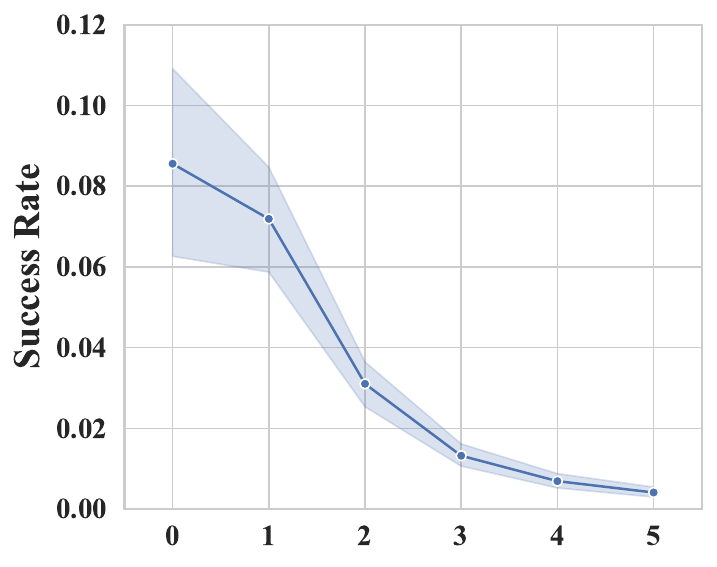}
		\caption{Auxiliary Variable Number}
		\label{fig:rq3:aux}
	\end{subfigure}
	\caption{Success rate of coefficient solving under different \flob\ parameter settings on ResNet18. In each subfigure, one parameter is fixed on the x-axis, while the other two vary. The curve shows the average success rate and the shaded area indicates the value range. Each solving attempt is given a timeout of 3,600 seconds, and timed-out cases are counted as failures.}
	\label{fig:rq3:success}
\end{figure*}

\subsubsection{Processing Time}\label{sec:rq3:performance}
\begin{figure}[t]
	\centering
	\includegraphics[scale=.5]{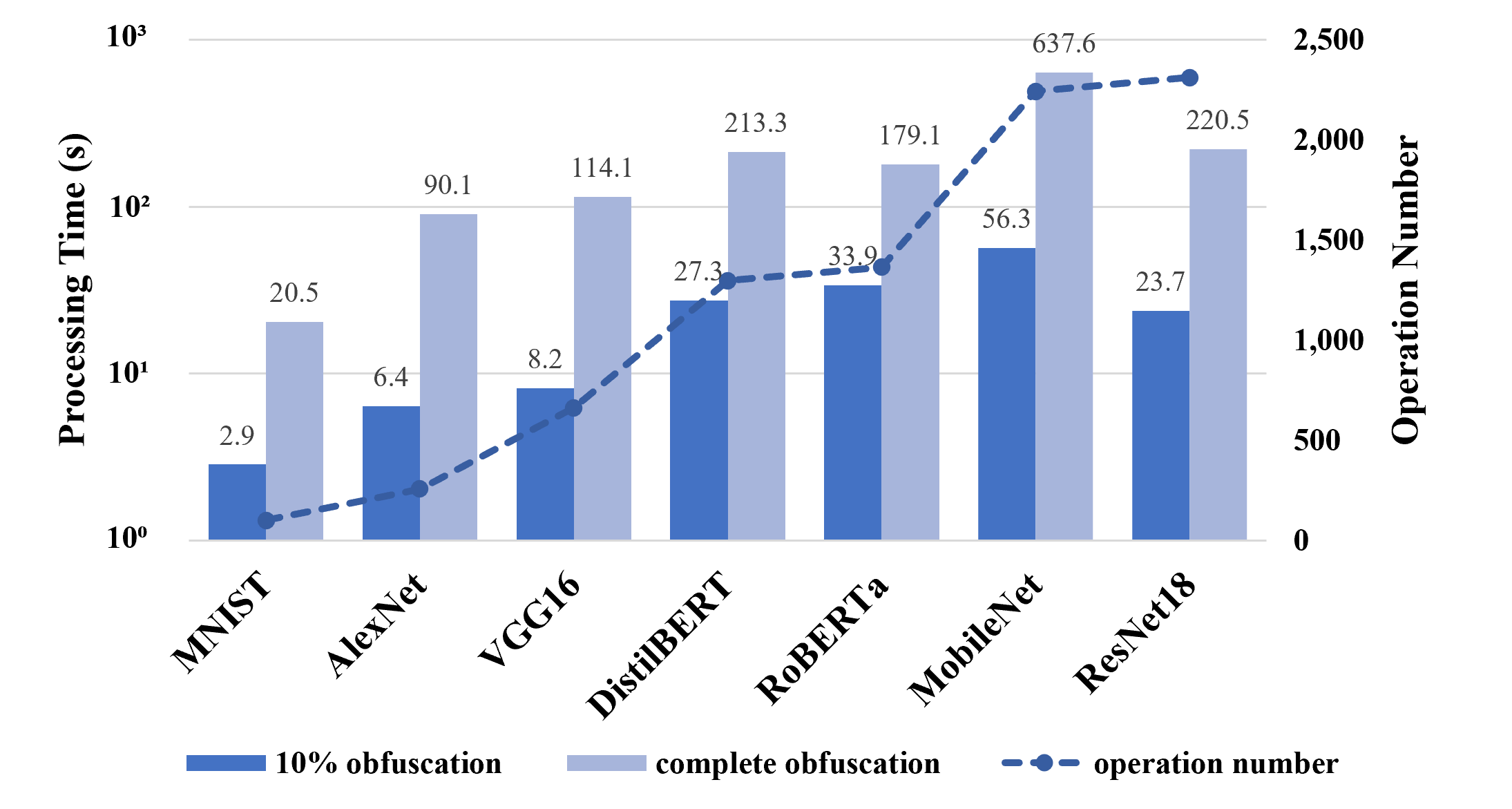}
	\caption{Obfuscation time of \flob. The left y-axis is shown in log scale.}
	\label{fig:rq3:time}
\end{figure}

The processing time of \flob\ is shown in Figure~\ref{fig:rq3:time}.
For each model, two settings are evaluated, namely obfuscating 10\% of the floating-point operations and all operations.
The dashed line simultaneously shows the number of floating-point operations in each model.
Overall, the processing time increases with the number of floating-point operations to be protected.
That is consistent with the design of \flob, i.e.,~more obfuscation instances leading to more term generation and coefficient solving during the process.

Figure~\ref{fig:rq3:success} further explains the major source of \flob’s processing cost.
In this experiment, ResNet18 is used as the target model, and the success rate of coefficient solving is evaluated under different \flob\ parameter settings.
Each solving attempt is given a timeout of 3,600 seconds, and timed-out cases are counted as failures.
In practice, most of the time is spent in coefficient solving, because terms are generated via sampling~(Algorithm~\ref{alg:term-gen}).
The sampled terms may not always yield a solvable coefficient system, and the generation-and-solving procedure must then be repeated until a valid solution is found~(Algorithm~\ref{alg:solving}).
Therefore, the overall processing time is heavily influenced by the success rate of coefficient solving.

Across all 432~($=9\times 8\times 6$)
parameter settings, the general trend is that the solving success rate decreases as the obfuscation parameters increase, i.e.,~as the obfuscation becomes stronger, indicating that stronger settings generally make coefficient solving harder, which in turn increases the total processing time.
Meanwhile, Figure~\ref{fig:rq3:term} shows a different trend, where the success rate reaches its peak when the number of terms is 4.
This pattern arises because the success of coefficient solving is affected by two competing effects.
When the number of terms is smaller than 4, increasing it provides more possible combinations among the generated terms, thereby increasing the chance of obtaining a solvable system.
Once the number of terms exceeds 4, however, the coefficient system becomes progressively harder to solve, and the success rate correspondingly declines.
The peak at 2 in Figure~\ref{fig:rq3:operator} follows the same intuition, as a small increase in the number of operators per term initially enriches the space of candidate terms and improves solvability, whereas further increases make the generated expressions more complex and reduce the solving success rate.

Although the solving success rate affects processing efficiency to some extent, this cost is largely incurred only once in the current design of \flob.
Successfully solved MBA expressions are stored in an expression database for later reuse~(as shown in Figure~\ref{fig:sys-overview}), which reduces repeated solving in subsequent obfuscation tasks and improves practical efficiency over time.

\subsubsection{Obfuscation Overhead}\label{sec:rq3:overhead}

\begin{figure*}[t]
	\centering
	\begin{subfigure}[t]{.32\textwidth}
		\centering
		\includegraphics[scale=.46]{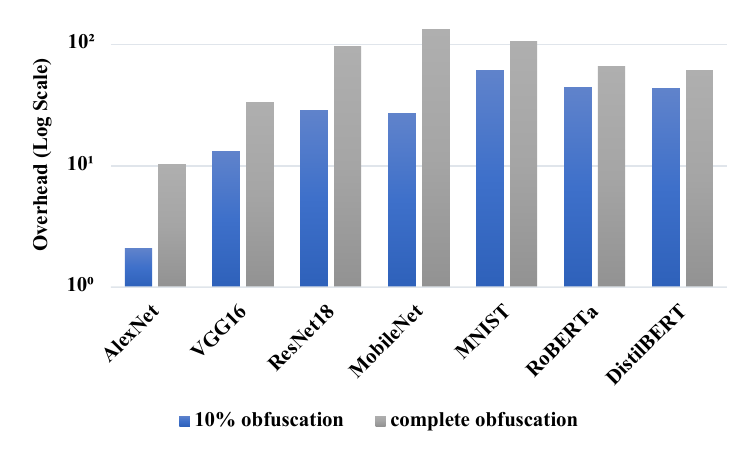}
		\caption{Execution Time}
		\label{fig:rq3:overhead:time}
	\end{subfigure}
	\hfill
	\begin{subfigure}[t]{.32\textwidth}
		\centering
		\includegraphics[scale=.46]{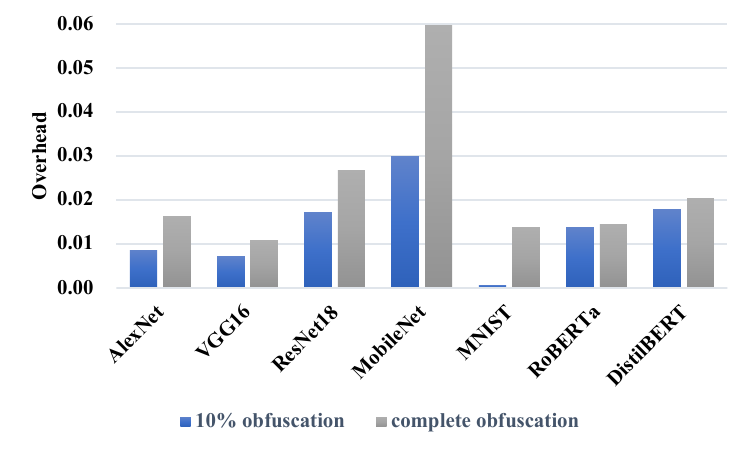}
		\caption{Runtime Memory}
		\label{fig:rq3:overhead:memory}
	\end{subfigure}
	\hfill
	\begin{subfigure}[t]{.32\textwidth}
		\centering
		\includegraphics[scale=.46]{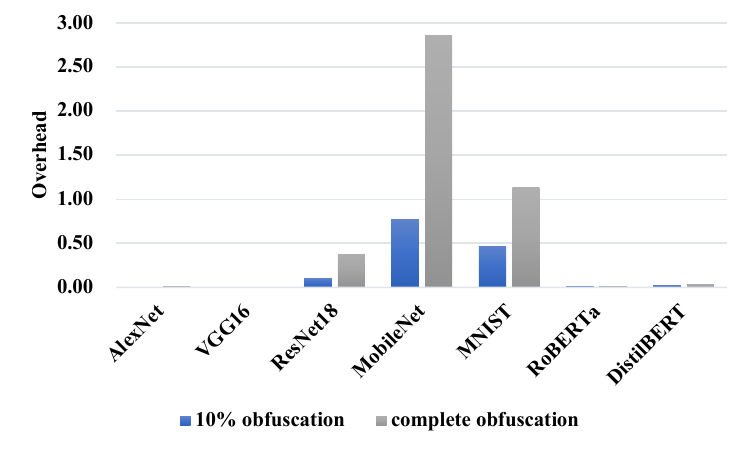}
		\caption{Binary Size}
		\label{fig:rq3:overhead:size}
	\end{subfigure}
	\caption{Overhead of \flob. The y-axis of execution time overhead is shown in log scale.}
	\label{fig:rq3:overhead}
\end{figure*}

The overhead introduced by \flob\ is shown in Figure~\ref{fig:rq3:overhead}, including execution time, runtime memory usage, and executable binary size under the settings of obfuscating 10\% of floating-point operations and all the operations, respectively.
Overall, stronger obfuscation consistently leads to higher overhead.

Execution time is the dominant overhead introduced by \flob, which is expected because its protection is achieved by deliberately complicating the original computation.
Under 10\% obfuscation, the execution-time overhead ranges from 2x to 62x, with an average of about 40x.
Under complete obfuscation, the range further increases to 10x to 133x, with an average of about 62x.
These results show a clear increase in runtime cost as more floating-point operations are protected.
At the same time, this overhead remains competitive when compared with traditional MBA-based obfuscation methods.
According to the \nnshield\ results~\cite{neuroshield}, \loki\ incurs an execution-time overhead of up to 1700x on MNIST, while \tigress\ reaches 194x-258x on MNIST, ResNet, and MobileNet.
By contrast, under complete obfuscation, the corresponding overhead of \flob\ on these models stays within 97x-134x, suggesting that it remains more efficient than prior MBA-oriented obfuscation tools in terms of execution-time overhead, while still providing strong protection.

In addition, runtime memory overhead and binary size overhead remain comparatively moderate. Across the evaluated models, the increase in runtime memory usage is generally small, while the binary-size overhead also stays within a limited range.
MobileNet is observed to exhibit larger memory and binary-size overhead than most other models. This may be related to its comparatively fine-grained and deeper model structure, which can produce more distributed obfuscation targets in the compiled binary, thereby introducing more transformed code and auxiliary data after obfuscation.

\begin{tcolorbox}[top=2.5pt, bottom=2.5pt, left=2.5pt, right=2.5pt]
	\textbf{Answer to RQ3}:
	\flob\ achieves practical performance with a controllable trade-off between protection strength and cost.
	Its execution-time overhead is lower than existing MBA-based obfuscation methods, while the runtime memory and binary-size overhead remain moderate.
\end{tcolorbox}
\section{Discussion and Future Work}


\subsection{General-Purpose Implementation}
The current prototype of \flob\ is implemented on top of TVM~\cite{tvm}, but the design itself is not tied to TVM.
In particular, TVM is used to obtain the LLVM IR of target models.
The core design of \flob, however, is platform-independent.
It operates on floating-point computations at the IR level and is not tied to a specific front-end framework or compiler stack.
Therefore, it is applicable to other compiler toolchains that expose comparable IR representations.


\subsection{Comparison with Related Work}\label{sec:disc:compare}
\mdobf~\cite{modelobfuscator} is similar to \nnshield, as protection is largely achieved by complicating code structure and integer-oriented computation. 
Since \flob\ already shows stronger resistance to reverse engineering than \nnshield~(\S\ref{sec:eva:rq2:re}), it can be inferred that \flob\ is unlikely to be weaker than \mdobf\ in this aspect.
\dynamo~\cite{dynamo} is not naturally applicable to all nonlinear operations.
By contrast, \flob\ directly protects floating-point computation at the IR level, giving it a broader applicability across operator types and deployment scenarios.
\nnscope~\cite{neuroscope} relies on precise identification of memory regions and access patterns to recover information from DNN binaries, whereas the increased complexity of memory accesses introduced by \flob\ would make such analysis less effective.

\subsection{Efficiency and Optimization}
The evaluation shows that \flob\ incurs noticeable runtime overhead, which is expected because protection is achieved by transforming original floating-point computations into more complex obfuscated forms.
Improving efficiency is therefore an important direction for future work.
One possibility is to reduce the precision used in the lifted integer representation.
While the current design reserves sufficient precision to avoid additional numerical error, DNN inference often tolerates limited numerical inaccuracy without affecting final prediction behavior, suggesting room for a more efficient accuracy-efficiency trade-off.
Another direction is to develop obfuscation methods for floating-point vector operations directly. The current reduction of multiplication and division partially disrupts the original parallelism of vectorized computation, and a vector-aware design may reduce this cost.

\subsection{Quantification of Obfuscation}
Conventionally, it is an open problem to quantify the strength of software obfuscation, because protection usually targets control flow and data flow, whose resistance is hard to measure precisely.
In contrast, \flob\ targets floating-point operations through MBA-based transformations, which may enable a more structured notion of protection strength, for example in terms of added computation or transformed instructions.
The definition of such metrics and their connection to practical resistance against automated analysis remain unclear, and are left for future work.

\section{Related Work}

\textbf{Code Obfuscation.}
Software obfuscation techniques aim to protect software intellectual property 
and hinder reverse engineering by transforming program code while preserving its functionality. 
General obfuscation schemes include opaque predicates~\cite{opaque-predicates}, 
control flow flattening~(CFF)~\cite{control-flow-flattening}, virtualization~(VM)~\cite{virtualization, xvmp},
and Mixed Boolean-Arithmetic~(MBA)~\cite{mba-definition}. 
While frameworks like Loki~\cite{loki} leverage these techniques for general software protection—and others, 
such as Obelix~\cite{obelix}, utilize hardware-based trusted execution environments—these approaches often fall short 
when applied to Deep Neural Network (DNN) binaries. 
Opaque predicates and control flow flattening are less effective for DNN binaries 
as they have simple control flow and are dominated by linear algebra computations. 
Virtualization-based obfuscation can introduce significant overhead for DNN inference, 
while MBA-based obfuscation can be easily simplified by existing deobfuscation techniques, 
such as MBA-Blast~\cite{mba-blast}.

\textbf{Code Deobfuscation.}
Recent deobfuscation efforts tackle complex software protections through structural and binary restoration. 
For instance, DeFFai~\cite{deffai} mitigates Control Flow Flattening~(CFF) via Abstract Interpretation, 
while QSynth~\cite{david2020qsynth} combines Dynamic Symbolic Execution~(DSE) and program synthesis to recover binary semantics. 
However, MBA obfuscation remains a primary bottleneck. 
To overcome the high performance overhead of early bit-blasting tools like MBA-Blast~\cite{mba-blast}, 
advanced white-box solutions have emerged: 
SiMBA~\cite{simba} and GaMBA~\cite{gamba} employ algebraic rewriting to simplify MBAs without complete bit-level decomposition, 
while ProMBA~\cite{promba} integrates program synthesis with term rewriting. 
Additionally, MBA-Solver~\cite{xu2021boosting} utilizes truth-table-derived signature vectors to safely reduce bitwise-arithmetic mixing. 
To overcome the exponential memory scaling issues of such table-based methods on multi-variable expressions, 
MBA-Flatten~\cite{mba-flatten} introduces an in-place transformation that directly unifies bitwise operations into an arithmetic form 
without requiring precomputed tables. 
To bypass the syntactic limitations of these white-box methods, novel paradigms have shifted towards black-box and AI-driven approaches. 
XSmir~\cite{xsmir} utilizes I/O-guided program synthesis with local inference rules to simplify custom MBAs, 
while NeuReduce~\cite{feng-etal-2020-neureduce} frames MBA simplification as a data-driven "machine translation" task using a Seq2Seq model, 
drastically accelerating deobfuscation over traditional symbolic solvers.

\textbf{DNN Protection.} 
DNN protection has attracted significant attention as models are increasingly deployed in untrusted environments. 
Prior work can be broadly categorized into three directions:
\begin{enumerate*}[label=\roman*)]

\item Cryptographic methods, including homomorphic encryption, garbled circuits, 
and two-party computation~\cite{secureml, he-sgx, bayhenn, minionn}; 
\item Hardware-based methods, which mainly apply trusted execution environments~\cite{tbnet, telekine, slalom}; 
\item Structure-targeted methods, including operator and insertion, computation reordering, control flow flattening,
and layer injection~\cite{neuroshield, modelobfuscator, neurobfuscator, zhang2025flatd};
\item Watermarking methods, which embed watermarks into model parameters or activations for ownership verification~\cite{deepsigns, themis}.
\end{enumerate*}
These efforts collectively highlight the trade-offs between security guarantees, computational overhead, 
and model utility in safeguarding DNN assets.

\textbf{DNN Binary Reverse Engineering.}
Reverse engineering threats against deployed DNNs have rapidly escalated 
from software decompilation to physical hardware exploitation. 
At the software level, tools like BTD~\cite{btd} and DnD~\cite{dnd} rely on static or dynamic analysis 
to decompile x86 and cross-architecture DNN binaries, recovering high-level model semantics. 
To bypass the heavy overhead of static analysis on IoT edge devices, 
NeuroScope~\cite{neuroscope} leverages dynamic memory tracing for efficient model extraction. 
Beyond software, system-level and physical attacks present severe risks: 
the Hermes Attack~\cite{hermes-attack} achieves lossless extraction by intercepting unencrypted PCIe traffic, 
while power side-channel analysis~\cite{side-channel-assisted-reverse-engineering} can reverse-engineer 
even closed-source, encrypted hardware accelerators (e.g., FPGAs) and their executing parameters. 
Together, these vectors thoroughly compromise model confidentiality and intellectual property.
\section{Conclusion}

We propose \flob, a DNN obfuscation framework that protects floating-point computations in deployed binaries.
By lifting floating-point values into a higher-precision representation domain, it enables MBA-based obfuscation while preserving the original computation semantics.
It further supports multiplication and division through operation reduction, allowing a wider range of floating-point operations to be protected.
The evaluation shows that \flob\ preserves exact correctness without introducing additional numerical error, while effectively reducing the success of existing MBA deobfuscation and DNN reverse-engineering methods. 
In addition, it achieves a practical trade-off between protection strength and efficiency, with lower execution-time overhead than existing MBA-based obfuscation methods.


\bibliographystyle{ACM-Reference-Format}
\bibliography{references}




\end{document}